 \documentclass[final,5p,times,twocolumn]{elsarticle}
\usepackage{amssymb}
\usepackage{amsmath}

\usepackage[colorlinks=true]{hyperref}
\usepackage{bm}
\usepackage{xcolor}
\usepackage{soul}
\usepackage{enumitem}

\journal{arXiv }

\begin{document}

\begin{frontmatter}

 \title{Ballistic surfing acceleration:  A physical alternative to diffusive shock acceleration for cosmic rays}

\author[label1]{Krzysztof Stasiewicz} 
\affiliation[label1]{organization={Space Research Centre, Polish Academy of Sciences},
            addressline={Bartycka 18A}, 
            city={Warszawa},
            postcode={00-716}, 
            country={Poland}}
 \fntext[label1]{Corresponding author:  kstasiewicz@cbk.waw.pl}
\begin{abstract}
Particle acceleration at collisionless shocks is widely invoked to explain the origin of cosmic rays and non-thermal emission throughout the Universe, yet its physical basis remains unresolved. While the standard paradigm of diffusive shock acceleration (DSA) attributes energization to stochastic shock crossings, its theoretical foundations have recently been called into question. A first-principles electrodynamic analysis instead identifies stochastic wave energization (SWE) and ballistic surfing acceleration (BSA) as the dominant mechanisms that accelerate particles from thermal to relativistic velocities at shocks. While SWE operates within the shock structure via strong wave-particle interactions, BSA is initiated once the particle gyroradius exceeds the shock transition width, operating primarily in the region external to the shock layer. Driven by the convective electric field, BSA quantitatively reproduces the observed cosmic-ray spectral index of $-2.7$ below the knee and $-3$ above the knee. The knee energy at 5 PeV is interpreted as a geometric condition occurring when the particle gyrodiameter matches the characteristic shock length ($\sim 1\,\mathrm{pc}$). This mechanism predicts acceleration times of approximately 500 years, consistent with astrophysical constraints, and requires potential drops of only $\sim 10\,\mathrm{TV}$ along supernova remnant shocks to account for the PeV-scale spectrum. When applied to relativistic electrons in galaxy-cluster merger shocks, BSA explains acceleration to Lorentz factors of $\gamma \sim 10^4$--$10^5$ and accounts for the observed properties of radio-relic emission. By grounding shock acceleration directly in electric-field work, BSA provides a physically sound framework that fundamentally challenges the validity of the diffusive paradigm, which rests upon unphysical assumptions.
\end{abstract}

\end{frontmatter}

\section{Introduction} 
The standard theoretical framework invoked to explain the observed spectrum of cosmic rays, as well as the acceleration of highly relativistic electrons responsible for synchrotron radiation in astrophysical environments, is diffusive shock acceleration (DSA)  \cite[e.g.,][]{Bell:1978, Drury:1983, Blandford:1987, Helder:2012}. DSA remains central to the interpretation of particle acceleration across a wide range of astrophysical settings   \cite[e.g.,][]{Perri2022}), including electron acceleration associated with non-thermal emission from supernova remnants and large-scale cluster merger shocks   \cite[e.g.,][]{vanWeeren2019, Giuffrida2022}. In these models, particles gain energy through repeated shock crossings mediated by reflections from unspecified ``scattering centers''.

The quantitative predictions of DSA are based on the so-called first-order Fermi acceleration formula \cite{Bell:1978}
\begin{equation}
\Delta \mathcal{E}_{\mathrm{DSA}}\;=\;\frac{4}{3}\frac{(V_\mathrm{u}-V_\mathrm{d})}{c}\mathcal{E}\;=\;\frac{4}{3}(1-c_n^{-1})\frac{V_\mathrm{u}}{c}\mathcal{E},
\label{eq:DSA}
\end{equation}
where $V_\mathrm{u}$ and $V_\mathrm{d}$ are the upstream and downstream bulk velocities, respectively, $c$ is the speed of light, and $c_n \equiv n_\mathrm{d}/n_\mathrm{u} = V_\mathrm{u}/V_\mathrm{d}$ is the density compression ratio, as implied by the continuity equation, $nV_x=\mathrm{const}$. Here, $\Delta \mathcal{E}_{\mathrm{DSA}}$ denotes the average energy gain of an initially isotropic particle distribution after completing one full shock-crossing cycle.

Within this framework, energy gain is dictated not by the specific electrodynamic properties of the scattering process, but solely by the kinematic velocity contrast between the upstream and downstream bulk flows. Despite omitting explicit electric fields, this kinematic formulation remains the cornerstone of the DSA paradigm. When paired with the probability $P \approx 1 - 4 V_\mathrm{d}/c$ that a particle remains in the shock vicinity after one cycle, it yields the canonical cosmic-ray energy spectrum \cite{Bell:1978}:
\begin{equation}
\frac{\mathrm{d}N}{\mathrm{d}\mathcal{E}} \propto \mathcal{E}^{s}, \quad
s = -\left( \frac{3}{c_n - 1} + 1 \right),
\label{eq:DSAS}
\end{equation}
which, for the standard compression ratio $c_n = 4$, yields a spectral index of $s = -2$. This value significantly deviates from the observed index of $s \approx -2.7$ and, notably, is entirely independent of the shock angle.

The core challenge to the DSA model is that Equation~(\ref{eq:DSA}) is demonstrably inconsistent with the fundamental principles of electrodynamics \cite{Stasiewicz:2025P}. Given that the DSA framework is ubiquitous across thousands of publications and serves as the prevailing explanation for relativistic particle acceleration, a rigorous reassessment of its physical foundations is now imperative. If the underlying energy-gain mechanism lacks a robust electrodynamic basis, then the resulting spectral index---and the vast theoretical edifice of astrophysical applications built upon it---requires systematic reevaluation.

In the next section, a physically grounded model of shock acceleration is presented, validated using high-resolution measurements at Earth's bow shock \cite{Stasiewicz:2023MN2}. As will be shown, this formulation naturally identifies the physical source of particle energization and avoids the conceptual difficulties inherent in DSA. The limitations of the DSA framework become apparent when the two approaches are compared on the basis of fundamental electrodynamics.

\section{Physical processes in shocks}
The trajectories of particles with rest mass $m_0$, velocity $\bm{v}$, and charge $q$ in electromagnetic fields $\bm{E}$ and $\bm{B}$ are governed by the relativistic momentum equation:
\begin{equation}
\frac{\mathrm{d}\bm{p}}{\mathrm{d}t} = q(\bm{E} + \bm{v}\times \bm{B}),
\label{eq:MM}
\end{equation}
where $\bm{p} = \gamma m_0 \bm{v}$ and $\gamma = (1 - v^2/c^2)^{-1/2}$.

The kinetic energy gain of a particle, as implied by Eq.~(\ref{eq:MM}), is determined entirely by the electric field:
\begin{equation}
\Delta \mathcal{E} = q \int \bm{E} \cdot \bm{v}\, dt,
\label{eq:DE}
\end{equation}
where the kinetic energy is
\begin{equation}
\mathcal{E} = ( p^2 c^2+m_0^2 c^4 )^{1/2} - m_0 c^2,
\end{equation}
while the total energy is $\mathcal{T}=( p^2 c^2+m_0^2 c^4 )^{1/2}=\gamma m_0c^2$.

Thus, any viable acceleration mechanism must be traceable to a well-defined electric field, including the inductive field, ${\nabla}{\times}\bm{E} = -\partial \bm{B} / \partial t$. Magnetic turbulence alone cannot produce net particle energization, since the magnetic part of the Lorentz force performs no work.

\begin{figure}[t]
\centering
\includegraphics[width=8.5cm]{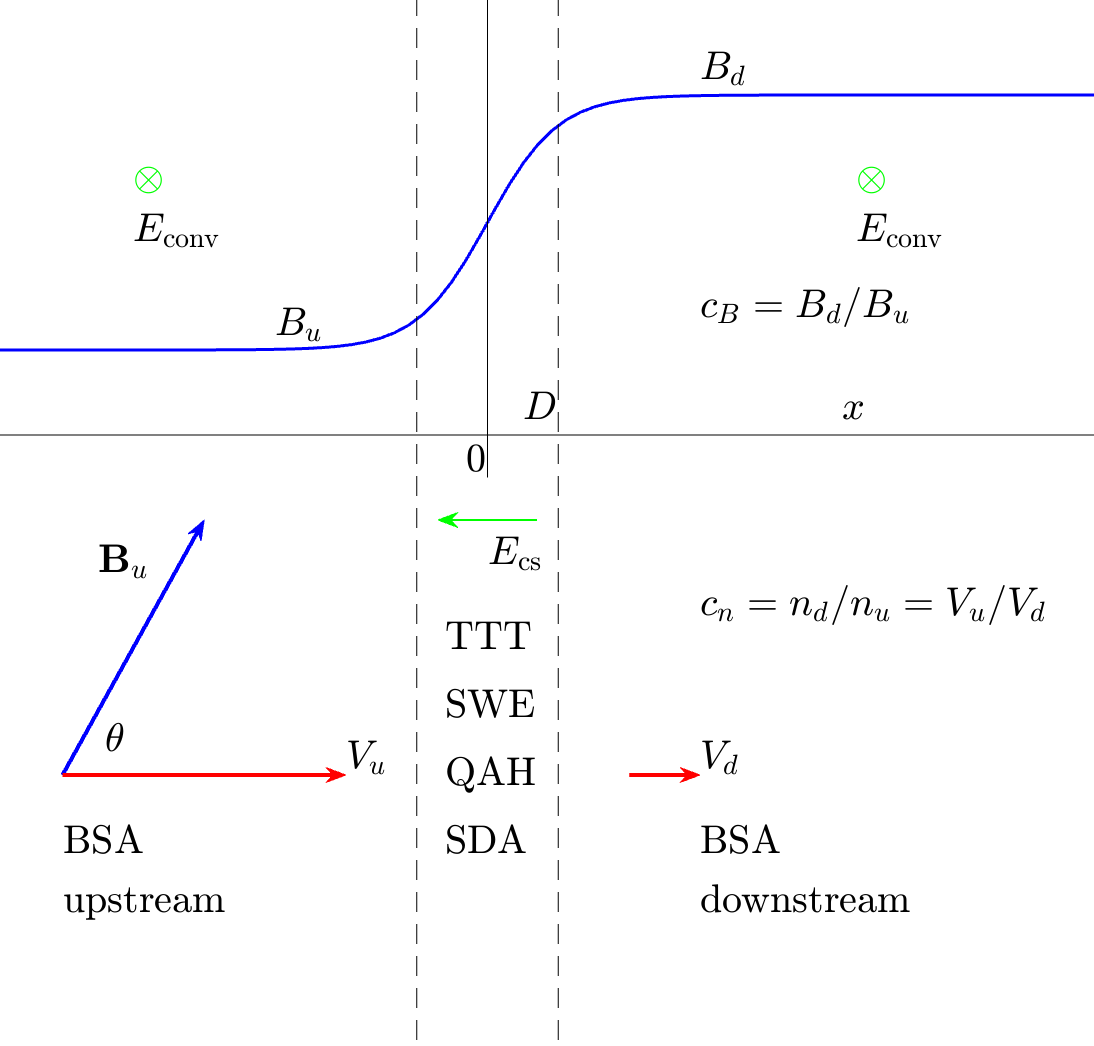}
\caption{Schematic structure of a collisionless shock in the normal incidence frame (NIF) showing the magnetic-field profile ($B$), electric fields ($E_\mathrm{conv},\, E_\mathrm{cs}$), and the plasma flows normal to the shock ($V_\mathrm{u}, V_\mathrm{d}$). The characteristic shock-transition scale, $D$,  is  of the order of the  thermal proton  gyroradius. Acceleration mechanisms operating at collisionless shocks are indicated: TTT (transit-time thermalization), SWE (stochastic wave energization), QAH (quasi-adiabatic heating), SDA (shock drift acceleration), and BSA (ballistic surfing acceleration). These processes provide a complete pathway for accelerating particles from thermal energies to relativistic regimes. \label{fig:1}}
\end{figure}

Consider the shock schematic illustrated in Figure~\ref{fig:1}, presented in the normal incidence frame (NIF) commonly adopted in numerical simulations. The shock is sustained by a plasma flow with upstream and downstream velocities $V_\mathrm{u}$ and $V_\mathrm{d}$, respectively, directed along the $+x$ axis. The magnetic-field profile is modeled as a ramp of the form $B_z(x) \propto \tanh(x/D)$, where $D$ denotes the characteristic half-width of the transition layer (the magnetic ramp), and $c_B = B_\mathrm{d}/B_\mathrm{u}$ represents the magnetic-field compression ratio.

In the shock  frame there is a convective electric field 
\begin{equation}
\bm{E}_{\rm conv} = -\bm{V} \times \bm{B}= \hat{\bm{y}} \,V_\mathrm{u} B_\mathrm{u}\sin\theta, 
\end{equation}
constant across a one-dimensional structure, as implied by  Maxwell's equations. 
Here, $\theta$ is the shock angle and $\hat{\bm{y}}$ is the unit vector in the shock tangential plane.
The  convective field is typically $E_\mathrm{conv}\approx 0.5\text{--}5\,\mathrm{mV\, m^{-1}}$ upstream of the Earth's bow shock.

The electron pressure gradient within the shock ramp produces a macroscopic cross-shock electric field, as implied by the generalized Ohm’s law:
\begin{equation}
\bm{E}_\mathrm{cs} = -\frac{\nabla P_\mathrm{e}}{e n},
\end{equation}
which is directed along the $-x$ axis.    In addition, various plasma instabilities generate strong electromagnetic and electrostatic turbulence within the shock, represented by fluctuating fields $\tilde{\bm{E}}(\bm{r},t),\,\tilde{\bm{B}}(\bm{r},t),\,{n}(\bm{r},t)$, which may extend both upstream and downstream as long-wavelength modes. The turbulence is primarily driven by cross-field current instabilities, with power spectral density maximized in the frequency range extending from below the proton cyclotron frequency ($\omega_\mathrm{ci}$) to the lower-hybrid frequency. However, the electric field wave amplitude increases significantly with frequency, reaching its peak ($\tilde{E} \sim 100\,\mathrm{mV\, m^{-1}}$) toward the electron cyclotron and ion plasma frequencies, as measured at the Earth's bow shock \cite{Stasiewicz:2021MN}.

Consequently, three distinct electric field components contribute to particle energization at shocks:
\begin{equation}
\bm{E} = \hat{\bm{y}} E_{\rm conv} + \hat{\bm{x}} E_{\rm cs}(x) + \tilde{\bm{E}}(\bm{r},t). 
\label{eq:E}
\end{equation}
By substituting Eq.~(\ref{eq:E}) into the work-energy relation (\ref{eq:DE}), the particle energization within the shock environment,  $|x|\le D$, can be expressed as:
\begin{align}
\Delta \mathcal{E}_{\mathrm{inside}} &= \int_{|x|\le D}^{\mathrm{SDA}} qE_\mathrm{conv} v_y \, dt + \int_{|x|\le D}^{\mathrm{CS}} qE_\mathrm{cs} v_x \, dt \nonumber \\
&\quad + \int_{|x|\le D}^{\mathrm{SWE}} q\tilde{\bm{E}} \cdot \bm{v} \, dt
\label{eq:inside}
\end{align}
where the first term corresponds to general shock drift acceleration (SDA), the second to acceleration by the cross-shock (CS) electric field, and the third to stochastic wave energization (SWE)---along with other wave-driven mechanisms, including the inductive electric field arising from $\partial \bm{B}/\partial t$.

Transit-time thermalization (TTT) is a fundamental, yet often overlooked, mechanism in which a significant fraction of the upstream kinetic energy, $m_\mathrm{i} V_\mathrm{u}^2 / 2$, is converted into ion gyration energy via kinematic interactions with the magnetic field gradient, independent of plasma turbulence \cite{Stasiewicz:2023MN2}. These test-particle simulations reveal that the incoming upstream ions are initially heated by TTT and stochastic wave energization (SWE), the latter being driven by strong wave electric fields ($\tilde{E} \approx 10\text{--}100\,\mathrm{mV\,m^{-1}}$) observed at shocks \cite{Stasiewicz:2023MN2, Stasiewicz:2023MN1,   Stasiewicz:2023MN3}.

SWE requires sufficiently steep electric-field gradients and can be driven by even a single coherent wave \cite{Stasiewicz:2021MN, Karney:1979, Stasiewicz:2000, Stasiewicz:2020AJ1}, resulting in deterministic chaos and divergent particle trajectories. In the presence of multiple wave modes or turbulence, chaos and Hamiltonian stochasticity are further enhanced by wave-particle resonance overlap. The velocity kick for a single encounter with a wave packet is determined by the wave electric field, yielding
\begin{equation}
\Delta v \lesssim \frac{\tilde{E}}{B}.
\label{eq:SWE}
\end{equation}
This acceleration is a resonant interaction \cite{Stasiewicz:2021MN, Stasiewicz:2022AG} characterized by a phase-favorable coupling between the wave electric field and the particle velocity, typically occurring once per gyroperiod.  As shown in Fig.~7 of \cite{Stasiewicz:2022AG}, acceleration takes place within brief, stochastically occurring coherence intervals. A limited number of such encounters during a rapid transit through quasi-perpendicular shocks can account for proton energies restricted to $\sim 20\,\mathrm{keV}$ \cite{Stasiewicz:2023MN2}.  Conversely, extended interaction zone and more numerous encounters in quasi-parallel shocks can accelerate protons to significantly higher energies of $\sim 200\,\mathrm{keV}$, as observed at Earth's bow shock \cite{Stasiewicz:2021JR, Stasiewicz:2022MN}.

The threshold for the onset of SWE is governed by a dimensionless function related to charge non-neutrality,   $\delta n_\mathrm{e} = n_\mathrm{i} - n_\mathrm{e}$  \cite{Stasiewicz:2020MN}:
\begin{equation}
\chi(\bm{r},t) = \frac{m\, \nabla \cdot \bm{E}_\perp}{q B^2} \approx \frac{m\, \delta n_\mathrm{e}}{Z\epsilon_0 B^2}, \qquad |\chi | \gtrsim 1, \label{eq:chi}
\end{equation}
where  $Z=q/e$, and  the divergence is expressed as $\nabla \cdot \bm{E}_\perp \approx \nabla \cdot \bm{E} = \delta n_\mathrm{e}\, e/\epsilon_0$, which for plane waves or 1D structures reduces to $\partial E_x/\partial x$. 
These properties of SWE have recently been fully corroborated by particle-in-cell (PIC) simulations \cite{Ha:2026AA}. Specifically, regions where $|\chi | > 1$ are identified as localized ion acceleration centers---physical structures that remain undefined within the traditional diffusive shock acceleration (DSA) framework.

Low-energy particles with gyroradii $r_\mathrm{g} \equiv p_\perp/qB$ smaller than the shock width ($r_\mathrm{g} < D$) undergo shock drift acceleration (SDA), which is equivalent to quasi-adiabatic heating (QAH) \cite{Stasiewicz:2023MN1}. While this process is observed for electrons, both SDA and QAH are suppressed for ions within the turbulent shock environment due to the violation of the first adiabatic invariant.

The concept of stochastic shock drift acceleration (SSDA), introduced in recent studies \cite[e.g.,][]{Katou:2019, Amano:2020, Kobzar:2021, Amano:2022} to explain the pre-acceleration of electrons, represents a contradiction in terms. This is because SDA relies on adiabatic particle drift motion along the convective electric field, whereas stochasticity arises from turbulence, which destroys adiabatic invariants and prevents regular drift motion and hence systematic energization by the convective electric field.

Due to the mass dependence in Eq.~(\ref{eq:chi}), electron stochasticity ($\chi_\mathrm{e} = \chi_\mathrm{i}/1836$) is much smaller than that of protons in the same turbulence, explaining their resistance to non-adiabatic (i.e., stochastic) heating under typical conditions at Earth's bow shock \cite{Stasiewicz:2023MN1}. Whereas terrestrial bow-shock turbulence typically leaves electrons in the quasi-adiabatic regime ($\chi_\mathrm{e}\lesssim 1$), turbulence at astrophysical shocks can be substantially stronger, enabling effective electron pre-acceleration by SWE when $\chi_\mathrm{e}>1$.

Once the gyroradius of SWE-accelerated particles exceeds the shock width, BSA becomes an increasingly important energization mechanism operating outside the shock
\cite{Stasiewicz:2025P, Stasiewicz:2023MN2} 
  \begin{equation}
 \Delta \mathcal{E}_{\mathrm{outside}}= \int_{|x|> D}^{\mathrm{BSA}} qE_\mathrm{conv} v_y\, dt + \int_{|x|> D}^{\mathrm{WE}} q\tilde{\bm{E}} \cdot \bm{v}\,dt. \label{eq:outside}
 \end{equation}
Wave energization (WE) outside shocks is constrained by the local wave potential  ($ \Delta \mathcal{E}\lesssim q\tilde{E}\lambda/2$) or the square of the phase speed ($ \Delta \mathcal{E}\lesssim mv^2_\mathrm{ph}/2$); it yields no significant net energy gain when averaged over timescales exceeding the wave period, as these waves remain below the stochastic wave energization (SWE) threshold. Consequently, while such waves do not contribute to cumulative acceleration, they provide the pitch-angle scattering necessary to facilitate shock-return trajectories and repetitive gyrations around the shock front, which in turn enable BSA.  The absence of any terms attributable to Fermi acceleration or DSA in general energization Equations (\ref{eq:inside}) and (\ref{eq:outside}) implies that these mechanisms lack a formal basis in first-principles electrodynamics.

Generally, the work done by the cross-shock electric field on a particle integrates to zero over a full gyration, as the energy gained in one direction is offset by an equivalent loss in the opposite direction. In contrast, the energy gain in BSA arises from the disparity between the upstream and downstream gyroradii, which ensures that the energy gain in the upstream region exceeds the energy loss in the downstream region, leading to a net dc acceleration:
\begin{equation}
\Delta \mathcal{E}\; \lesssim \;2(r_\mathrm{gu} - r_\mathrm{gd})qE_\mathrm{conv}= 2(1-c_B^{-1})r_\mathrm{gu}qE_\mathrm{conv}. \label{eq:DEB}
\end{equation}
This expression applies to particles with large gyroradii at normal incidence to quasi-perpendicular shocks (see Fig.~3 of \cite{Stasiewicz:2025P}). Because this BSA process operates independently of turbulence, it is an omnipresent feature of quasi-perpendicular shocks that is entirely absent from the standard DSA formulation and other alternative shock acceleration models. Omitting such a dominant energization process renders the DSA framework physically incomplete.

\section{Cosmic-ray acceleration by BSA}
Consider a cosmic-ray proton with kinetic energy $\mathcal{E} = 1$\,GeV, near the onset of the observed power-law spectrum. It experiences only $\sim$1\,ms of interaction within a shock of width $2D = 300$\,km, whereas it undergoes approximately 70\,s of continuous ballistic surfing acceleration by the convective electric field in the upstream region, where its gyroradius is $r_\mathrm{g} \approx 6 \times 10^6$~km for $B_\mathrm{u} = 1$\,nT. This five-orders-of-magnitude disparity in interaction time demonstrates that for high-energy particles ($r_\mathrm{g} \gg D$), BSA operates as the primary acceleration mechanism, rendering the contributions of internal shock processes (TTT, SDA/QAH, CS, WE, SWE) and $\Delta \mathcal{E}_{\mathrm{inside}}$ negligible.

Averaging the energy gain in Eq.~(\ref{eq:DEB}) over both incident and pitch  angles for an arbitrary shock geometry yields \cite{Stasiewicz:2025P}:
\begin{equation}
\Delta \mathcal{E} \approx \Delta \mathcal{E}_{\mathrm{BSA}} = g \sin\theta\, (1 - c_B^{-1}) \frac{V_\mathrm{u}}{c} \mathcal{E},
\label{eq:BSA}
\end{equation}
where $g \lesssim 1$ is a geometric factor depending on the shock angle $\theta$ and the initial pitch-angle distribution, attaining its maximum value for perpendicular shocks ($\theta = 90^\circ$).  This expression represents a physically grounded alternative to Eq.~(\ref{eq:DSA}).

Applying the same derivation that links Eq.~(\ref{eq:DSA}) to Eq.~(\ref{eq:DSAS}), while replacing diffusive shock acceleration with the BSA mechanism of Eq.~(\ref{eq:BSA}), yields the cosmic-ray spectral index \cite{Stasiewicz:2025P},
\begin{equation}
s = -\left( \frac{4(g\sin\theta)^{-1}}{c_B - 1} + 1 \right),
\label{eq:BSAS}
\end{equation}
which supersedes the DSA prediction of Eq.~(\ref{eq:DSAS}).
Because the parameter $g \sin\theta$ depends on the local---and typically unknown---shock geometry and particle pitch-angle distribution, we combine the universally observed spectral index ($s \approx -2.7$) with the canonical compression ratio ($c_B = 4$) to infer an effective value of $g \sin\theta \approx 0.78$.  Any spectral modifications arising from cosmic-ray propagation effects can be accommodated directly by adjusting the source spectral index via $g \sin\theta$.

Numerous simulation studies  \cite[e.g.,][]{Caprioli:2014a, Ha:2018,Kobzar:2021} successfully capture the collective shock processes responsible for the internal energization $\Delta \mathcal{E}_\mathrm{inside}$ described by Eq.~(\ref{eq:inside}). While these processes---specifically SWE \cite{Ha:2026AA}---provide the necessary injection velocity to satisfy the condition $r_\mathrm{g} > D$ required to initiate BSA, they play no role in determining the shape of the cosmic-ray spectrum. This is because $\Delta \mathcal{E}_\mathrm{BSA} \sim qr_\mathrm{gu}  E_\mathrm{conv} \gg \Delta \mathcal{E}_\mathrm{inside}$, and cosmic-ray acceleration is driven by ballistic surfing, occurring outside the shock transition and beyond both the spatial limits and typical temporal scales of these simulation boxes.

Using Eq.~(\ref{eq:BSA}) for relativistic particles with kinetic energy $\mathcal{E} \approx \gamma m_0 c^2$ and gyrofrequency $\omega_\mathrm{c} = qB/\gamma m_0$, the energy-gain rate can be expressed in terms of the Lorentz factor for particles gyrating around the shock:
\begin{equation}
\dot{\gamma}_{\rm BSA}
= \eta \frac{q V_\mathrm{u} B_\mathrm{u}}{m_0 c},
\qquad
\eta \equiv \frac{g \sin\theta}{4\pi}(c_B - c_B^{-1}).
\label{eq:gammadot_acc}
\end{equation}
It shows that the energy increases linearly with time through discrete gains per gyroperiod, and depends explicitly on the convective electric field through the identity $E_\mathrm{conv} \equiv V_\mathrm{u} B_\mathrm{u} \sin\theta$. For $g \sin\theta \approx 0.78$, as inferred from the cosmic-ray spectrum, and  compression $c_B \approx 4$ the efficiency factor is $\eta \approx 0.23$. 

Equation~(\ref{eq:gammadot_acc}) has been successfully applied in \cite{Ha2026} to account for the acceleration of electrons to Lorentz factors $\gamma \sim 10^4$–$10^5$, required for observable radio synchrotron emission from galaxy clusters \cite[][]{Hoeft2007, Siemieniec-Ozieblo2011, Pinzke2013, Stroe2016}. In contrast to standard DSA-based treatments, the acceleration rate follows directly  from the coherent shock electric field, without invoking phenomenological diffusion coefficients. The resulting steady-state electron spectra naturally reproduce the observed spectral curvature and high-frequency steepening of radio relics.

Equation~(\ref{eq:gammadot_acc}) implies that particles starting from $\sim$1~GeV can reach the cosmic-ray knee energy at $\mathcal{E}_{\rm knee}\approx 5$~PeV after about 230~years of net acceleration time in shocks with $V_\mathrm{u} = 10,000\,\mathrm{km\,s^{-1}}$ and $B_\mathrm{u} = 1\,\mathrm{nT}$, which corresponds to about 320 relativistic  gyroperiods for protons. At the knee energy the gyroradius is $r_\mathrm{g} \sim 1\,\mathrm{pc}$ for both protons and electrons. This result can be readily rescaled to other astrophysical environments by noting that the acceleration rate scales linearly with the product $V_\mathrm{u} B_\mathrm{u}$, whereas the gyroradius scales as $B^{-1}$.

The "knee" energy corresponds to the geometric condition $2r_\mathrm{gd} \approx L$, where $r_\mathrm{gd}$ is the downstream gyroradius and $L$ is the characteristic coherence length of the shock \cite{Stasiewicz:2025P}. This condition defines the knee energy as:
\begin{equation}
\mathcal{E}_{\rm knee} \approx \frac{Zec}{2} \langle B_\mathrm{d} L \rangle,
\label{eq:K}
\end{equation}
which yields $\mathcal{E}_{\rm knee} \approx 5\,\mathrm{PeV}$ for $\langle B_\mathrm{d} L \rangle \approx 1\,\mathrm{nT\,pc}$. This expression is independent of both the shock speed and the particle species, underscoring its universality as a fundamental physical relation. However, it maintains a critical dependence on the ion charge, a characteristic that provides a direct path for empirical validation.
The associated electric potential difference along the shock can be estimated as $\Phi_{\rm snr} \approx E_{\rm conv} L \approx \sin\theta \,V_\mathrm{u} B_\mathrm{u} L$. By substituting Eq.~(\ref{eq:K}), we find:
\begin{equation}
\Phi_{\rm snr} \approx \frac{2\sin\theta}{c_B} \frac{V_\mathrm{u}}{c}\, \frac{\mathcal{E}_{\rm knee}}{e} \approx 8 \times 10^{12}\,\mathrm{V}, 
\label{eq:V}
\end{equation}
which directly links the observed knee energy to the maximum potential available in supernova remnants during the expansion phase. The numerical value is computed for $c_B = 4$, $\theta \approx 90^\circ$, and $V_\mathrm{u} = 1000\,\mathrm{km\,s^{-1}}$. This result demonstrates that the PeV-scale cosmic-ray spectrum is generated by potentials of only several Teravolts.

Furthermore, these same shocks, characterized by potentials of $\sim 10\,\mathrm{TV}$, can also account for the cosmic-ray spectrum at energies above the knee. This occurs when the plasma flow gradually stagnates in the shock frame in the far-downstream region ($V_\mathrm{d} \approx 0$), causing the convective electric field to vanish. In this regime, gyrating particles undergo sustained acceleration in the upstream region without the energy losses   typically associated with downstream transits. Such conditions could potentially push particle energies toward the "ankle" of the cosmic-ray spectrum at 5\,EV. The observed steepening to a spectral index of $s \approx -3$ is then quantitatively explained by Eq.~(\ref{eq:BSAS}) with a geometric factor of $g\sin\theta = 0.66$.

\section{Physical inconsistency of DSA} 

In contrast to the electrodynamically grounded BSA model described above, the DSA framework rests upon unphysical premises. Specifically, it omits the universal energy gain in Eq.~(\ref{eq:DEB}) inherent to quasi-perpendicular shocks and features no accelerating electric field, whether explicit or implicit. Because a magnetic field can do no work on charged particles, the lack of an accelerating electric field renders the DSA framework electrodynamically incomplete and fundamentally unphysical. Additional deficiencies of the DSA framework are enumerated below:

\begin{enumerate}[label=\alph*)]
\item As Equation~(\ref{eq:DSA}) was derived in a standard NIF frame for parallel shocks ($\theta = 0^\circ$), it operates in a regime where the convective electric field vanishes ($E_{\mathrm{conv}} = 0$). Without the postulation of alternative fields, and in light of the energy equation~(\ref{eq:DE}), particle acceleration is \emph{a priori excluded} from the model. Consequently, the cornerstone of DSA is built upon a physical impossibility, invalidating the paradigm at its source.

\item Equation~(\ref{eq:DSA}) was derived from the relativistic energy transformations between the upstream and downstream inertial frames, as reflected in its dependence on the compression ratio $c_n = V_\mathrm{u}/V_\mathrm{d}$. This represents a fundamental conflation because a kinematic energy difference expressed in two reference frames does not constitute physical energization. To be valid, acceleration must be evaluated within a single reference frame using Eq.~(\ref{eq:DE}), a process that leads instead to the physically grounded Eq.~(\ref{eq:BSA}), rendering Eq.~(\ref{eq:DSA})  invalid.

\item Standard DSA models, as well as the original Fermi acceleration mechanism \cite{Fermi:1949}, rely on the heuristic that particles gain energy through elastic reflections from moving magnetic-field gradients at shock fronts or within electromagnetic turbulence. However, this assumption is fundamentally inconsistent with electrodynamics; magnetic gradients are incapable of reflecting high-energy particles once their gyroradii exceed the gradient scale length. As demonstrated in \cite{Stasiewicz:2025P}, such gradients facilitate pitch-angle scattering through transit-time thermalization (TTT)---a process that conserves total particle energy in the local plasma frame while merely redistributing it between the parallel and perpendicular degrees of freedom.

\item Equation~(\ref{eq:DSA}) relies on the assumption of symmetric energy gain during upstream-to-downstream and downstream-to-upstream shock transits. This symmetry, however, is fundamentally inconsistent with electrodynamics. By integrating the momentum equation, it has been demonstrated in \cite{Stasiewicz:2025P} that particles lose energy in the downstream region and gain energy solely by "surfing" on the upstream side. The shock transitions themselves do not contribute significantly to energy change, regardless of the crossing direction.

\item The DSA parameters present in Eq.~(\ref{eq:DSA}), such as the velocity contrast $(V_\mathrm{u} - V_\mathrm{d})$ and the density compression ratio $c_n$, neither appear in nor can be derived from the general energization expressions in Eqs.~(\ref{eq:inside}) and~(\ref{eq:outside}). This indicates that the DSA framework lacks a formal, rigorous derivation from first-principles electrodynamics.
\end{enumerate}

A superficial similarity between the numerical values produced by the physically consistent BSA model (\ref{eq:BSA}), (\ref{eq:BSAS}) and the DSA model (\ref{eq:DSA}), (\ref{eq:DSAS}) for quasi-perpendicular shocks with compression ratios of $3$--$4$ helps explain why the DSA paradigm has enjoyed widespread acceptance despite its flawed physical foundation. While plasma density is anti-correlated with magnetic field strength in slow-mode magnetosonic structures, these variations are positively correlated in fast-mode (supersonic) structures, implying $c_n \approx c_B$. Consequently, the dependence of the spectral index and particle energy gain on $c_n$ in Eqs.~(\ref{eq:DSA}) and~(\ref{eq:DSAS}) has been misinterpreted as a causal physical relationship; in reality, the governing parameter is $c_B$ in Eq.~(\ref{eq:BSA}).

\section{Conclusions}
This paper presents a concise and physically grounded description of particle acceleration at collisionless shocks. It is argued that stochastic wave energization (SWE) and ballistic surfing acceleration (BSA) are the dominant mechanisms that accelerate particles from thermal to relativistic velocities at shocks. While SWE operates within the shock structure via strong wave-particle interactions, BSA is initiated once the particle gyroradius exceeds the shock transition width and operates in the region external to the shock. BSA explains several properties of cosmic rays and radio relics in galaxy clusters:

\begin{enumerate}[label=\arabic*)]
\item The kinetic energy of cosmic-ray particles is supplied directly by the convective electric field via BSA---a process driven by the disparity between the upstream and downstream gyroradii at shocks  and quantified by Eq.~(\ref{eq:DEB}).

\item BSA reproduces the observed cosmic-ray spectral index of $s \approx -2.7$ for energies below the $5\,\mathrm{PeV}$ knee by utilizing the adjustable parameter $g\sin\theta \approx 0.78$. This parameter effectively accounts for the prevailing, yet generally unknown, shock geometry and particle pitch-angle distribution in supernova remnants.

\item BSA naturally explains the knee energy as the threshold at which the particle gyrodiameter becomes comparable to the characteristic coherence length of the shock. This condition corresponds to $\langle B_\mathrm{d} L\rangle \sim 1\,\mathrm{nT\,pc}$, requiring a potential drop of $\Phi_{\mathrm{snr}} \sim 10\,\mathrm{TV}$ sustained during the supernova remnant expansion.

\item The model predicts an acceleration timescale to reach the knee energy, starting from $\sim 1\,\mathrm{GeV}$, of approximately 230~years in shocks with $B \sim 1\,\mathrm{nT}$ and $V_\mathrm{u} \sim 10,000\,\mathrm{km\,s^{-1}}$, corresponding to about 320 relativistic gyroperiods for protons in orbits intersecting the shock.

\item BSA could also explain the acceleration of particles beyond the "knee." This occurs in regimes where the plasma flow becomes stagnant in the far-downstream region, causing the convective electric field to vanish. Consequently, gyrating particles are not decelerated during downstream transits, allowing for nearly monotonic energy gain during upstream surfing. For example, it would require only 50 additional gyroperiods to increase particle energy from $5\,\mathrm{PeV}$ to $5\,\mathrm{EeV}$ at a shock with $V_\mathrm{u}\approx 10,000\,\mathrm{km\,s^{-1}}$ and, according to Eq.~(\ref{eq:V}), a $100\,\mathrm{TV}$ potential.

\item When applied to the acceleration of relativistic electrons responsible for radio-relic emission in galaxy clusters, BSA quantitatively accounts for the observed properties of this radiation \cite{Ha2026}.
\end{enumerate}

Taken together, these results indicate that SWE and BSA constitute a physically grounded and quantitatively predictive chain of processes for shock heating and acceleration across a broad range of astrophysical environments. Their explicit grounding in electrodynamics, through direct energy transfer from the electric field, provides a transparent physical basis for particle energization and a coherent framework for understanding both cosmic-ray spectra and radio-relic emission.

In contrast, the analysis presented in the preceding section exposes fundamental conceptual and physical inconsistencies within the DSA formulation, wherein each of the enumerated arguments independently invalidates it as a physical mechanism. The DSA formalism excludes the dominant acceleration mechanism at shocks and conflates frame-dependent energy transformations with physical acceleration. The popularity of DSA, as expressed in many publications, may instead stem from circumstantial parameter correlations rather than a rigorous representation of the underlying physical processes.


\section*{Declaration of competing interest}
The authors declare that they have no competing interests.

\section*{Data/software availability}
No datasets were generated or analyzed in this study.

 \bibliographystyle{elsarticle-num} 


\begin{thebibliography}{10}
\expandafter\ifx\csname url\endcsname\relax
  \def\url#1{\texttt{#1}}\fi
\expandafter\ifx\csname urlprefix\endcsname\relax\def\urlprefix{URL }\fi
\expandafter\ifx\csname href\endcsname\relax
  \def\href#1#2{#2} \def\path#1{#1}\fi

\bibitem{Bell:1978}
A.~R. Bell, The acceleration of cosmic rays in shock fronts - {I}, MNRAS
  182~(2) (1978) 147--156.
\newblock \href {https://doi.org/10.1093/mnras/182.2.147}
  {\path{doi:10.1093/mnras/182.2.147}}.

\bibitem{Drury:1983}
L.~O. Drury, An introduction to the theory of diffusive shock acceleration of
  energetic particles in tenuous plasmas, Reports on Progress in Physics 46~(8)
  (1983) 973--1027.
\newblock \href {https://doi.org/10.1088/0034-4885/46/8/002}
  {\path{doi:10.1088/0034-4885/46/8/002}}.

\bibitem{Blandford:1987}
R.~Blandford, D.~Eichler, Particle acceleration at astrophysical shocks: A
  theory of cosmic ray origin, Physics Reports 154~(1) (1987) 1 -- 75.
\newblock \href {https://doi.org/10.1016/0370-1573(87)90134-7}
  {\path{doi:10.1016/0370-1573(87)90134-7}}.

\bibitem{Helder:2012}
E.~Helder, J.~Vink, A.~Bykov, Y.~Ohira, J.~Raymond, R.~Terrier, Observational
  signatures of particle acceleration in supernova remnants, Space Sci. Rev.
  173 (2012) 369–431.
\newblock \href {https://doi.org/10.1007/s11214-012-9919-8}
  {\path{doi:10.1007/s11214-012-9919-8}}.

\bibitem{Perri2022}
S.~Perri, A.~Bykov, H.~Fahr, H.~Fichtner, J.~Giacalone, Recent developments in
  particle acceleration at shocks: Theory and observations, Space Science
  Reviews 218 (2022) 26.
\newblock \href {https://doi.org/10.1007/s11214-022-00892-5}
  {\path{doi:10.1007/s11214-022-00892-5}}.

\bibitem{vanWeeren2019}
R.~J. {van Weeren}, F.~{de Gasperin}, H.~{Akamatsu}, M.~{Br{\"u}ggen},
  L.~{Feretti}, H.~{Kang}, A.~{Stroe}, F.~{Zandanel}, {Diffuse Radio Emission
  from Galaxy Clusters}, Space Science Reviews 215~(1) (2019) 16.
\newblock \href {https://doi.org/10.1007/s11214-019-0584-z}
  {\path{doi:10.1007/s11214-019-0584-z}}.

\bibitem{Giuffrida2022}
R.~Giuffrida, M.~Miceli, D.~Caprioli, A.~Decourchelle, J.~Vink, S.~Orlando,
  F.~Bocchino, E.~Greco, G.~Peres, The supernova remnant sn 1006 as a galactic
  particle accelerator, Nature Communications 13 (2022) 5098.
\newblock \href {https://doi.org/10.1038/s41467-022-32781-4}
  {\path{doi:10.1038/s41467-022-32781-4}}.

\bibitem{Stasiewicz:2025P}
K.~Stasiewicz, Reinterpretation of {F}ermi acceleration of cosmic rays in terms
  of ballistic surfing acceleration in supernova shocks, Physics 7~(4) (2025)
  51.
\newblock \href {https://doi.org/10.3390/physics7040051}
  {\path{doi:10.3390/physics7040051}}.

\bibitem{Stasiewicz:2023MN2}
K.~Stasiewicz, Transit time thermalization and the stochastic wave energization
  of ions in quasi-perpendicular shocks, MNRAS 524 (2023) L50--L54.
\newblock \href {https://doi.org/10.1093/mnrasl/slad071}
  {\path{doi:10.1093/mnrasl/slad071}}.

\bibitem{Stasiewicz:2021MN}
K.~Stasiewicz, B.~Eliasson, Ion acceleration to 100 {keV} by the {ExB} wave
  mechanism in collisionless shocks, MNRAS 508 (2021) 1888--1896.
\newblock \href {https://doi.org/10.1093/mnras/stab2739}
  {\path{doi:10.1093/mnras/stab2739}}.

\bibitem{Stasiewicz:2023MN1}
K.~Stasiewicz, B.~Eliasson, Electron heating mechanisms at the bow shock -
  revisited with {M}agnetospheric {M}ultiscale measurements, MNRAS 520 (2023)
  3238--3244.
\newblock \href {https://doi.org/10.1093/mnras/stad361}
  {\path{doi:10.1093/mnras/stad361}}.

\bibitem{Stasiewicz:2023MN3}
K.~Stasiewicz, Origin of flat-top electron distributions at the {E}arth’s bow
  shock, MNRAS 527 (2023) L71--L75.
\newblock \href {https://doi.org/10.1093/mnrasl/slad146}
  {\path{doi:10.1093/mnrasl/slad146}}.

\bibitem{Karney:1979}
C.~F. Karney, Stochastic ion heating by a lower hybrid wave, Phys. Fluids 22
  (1979) 2188.
\newblock \href {https://doi.org/10.1063/1.862512}
  {\path{doi:10.1063/1.862512}}.

\bibitem{Stasiewicz:2000}
K.~Stasiewicz, R.~Lundin, G.~Marklund, Stochastic ion heating by orbit
  chaotization on nonlinear waves and structures, Physica Scripta T84 (2000)
  60--63.
\newblock \href {https://doi.org/10.1238/physica.topical.084a00060}
  {\path{doi:10.1238/physica.topical.084a00060}}.

\bibitem{Stasiewicz:2020AJ1}
K.~Stasiewicz, B.~Eliasson, Quasi-adiabatic and stochastic heating and particle
  acceleration at quasi-perpendicular shocks, The Astrophysical Journal 903~(1)
  (2020) 57.
\newblock \href {https://doi.org/10.3847/1538-4357/abb825}
  {\path{doi:10.3847/1538-4357/abb825}}.

\bibitem{Stasiewicz:2022AG}
K.~Stasiewicz, Z.~K{\l}os, Fine structure and motion of the bow shock and
  particle energisation mechanisms inferred from magnetospheric multiscale
  ({MMS}) observations, Ann. Geophys. 40 (2022) 315--325.
\newblock \href {https://doi.org/10.5194/angeo-40-315-2022}
  {\path{doi:10.5194/angeo-40-315-2022}}.

\bibitem{Stasiewicz:2021JR}
K.~Stasiewicz, B.~Eliasson, I.~J. Cohen, D.~L. Turner, R.~E. Ergun, Local
  acceleration of protons to 100 {keV} by the {ExB} wave mechanism in a
  quasi-parallel bow shock, Journal of Geophysical Research: Space Physics 126
  (2021) e2021JA029477.
\newblock \href {https://doi.org/10.1029/2021JA029477}
  {\path{doi:10.1029/2021JA029477}}.

\bibitem{Stasiewicz:2022MN}
K.~Stasiewicz, Z.~K{\l}os, On the formation of quasi-parallel shocks, magnetic
  and electric field turbulence, and ion energisation mechanism, MNRAS 513
  (2022) 5892--5899.
\newblock \href {https://doi.org/10.1093/mnras/stac1193}
  {\path{doi:10.1093/mnras/stac1193}}.

\bibitem{Stasiewicz:2020MN}
K.~Stasiewicz, Stochastic ion and electron heating on drift instabilities at
  the bow shock, MNRAS 496 (2020) L133--L137.
\newblock \href {https://doi.org/10.1093/mnrasl/slaa090}
  {\path{doi:10.1093/mnrasl/slaa090}}.

\bibitem{Ha:2026AA}
J.-H. Ha, K.~Stasiewicz, Stochastic wave energization in collisionless shocks
  using {PIC} simulations, Astronomy and Astrophysics 711 (2026) A82.
\newblock \href {https://doi.org/10.1051/0004-6361/202660321}
  {\path{doi:10.1051/0004-6361/202660321}}.

\bibitem{Katou:2019}
T.~Katou, T.~Amano, Theory of stochastic shock drift acceleration for electrons
  in the shock transition region, The Astrophysical Journal 874~(2) (2019) 119.
\newblock \href {https://doi.org/10.3847/1538-4357/ab0d8a}
  {\path{doi:10.3847/1538-4357/ab0d8a}}.

\bibitem{Amano:2020}
T.~Amano, T.~Katou, N.~Kitamura, M.~Oka, Y.~Matsumoto, M.~Hoshino, Y.~Saito,
  S.~Yokota, B.~L. Giles, W.~R. Paterson, C.~T. Russell, O.~Le~Contel, R.~E.
  Ergun, P.-A. Lindqvist, D.~L. Turner, J.~F. Fennell, J.~B. Blake,
  Observational evidence for stochastic shock drift acceleration of electrons
  at the earth's bow shock, Phys. Rev. Lett. 124 (2020) 065101.
\newblock \href {https://doi.org/10.1103/PhysRevLett.124.065101}
  {\path{doi:10.1103/PhysRevLett.124.065101}}.

\bibitem{Kobzar:2021}
O.~Kobzar, J.~Niemiec, T.~Amano, M.~Hoshino, S.~Matsukiyo, Y.~Matsumoto,
  M.~Pohl, Electron acceleration at rippled low-mach-number shocks in high-beta
  collisionless cosmic plasmas, The Astrophysical Journal 919~(2) (2021) 97.
\newblock \href {https://doi.org/10.3847/1538-4357/ac1107}
  {\path{doi:10.3847/1538-4357/ac1107}}.

\bibitem{Amano:2022}
T.~Amano, M.~Hoshino, Theory of electron injection at oblique shock of finite
  thickness, The Astrophysical Journal 927~(1) (2022) 132.
\newblock \href {https://doi.org/10.3847/1538-4357/ac4f49}
  {\path{doi:10.3847/1538-4357/ac4f49}}.

\bibitem{Caprioli:2014a}
D.~Caprioli, A.~Spitkovsky, Simulations of ion acceleration at non-relativistic
  shocks. {I.} acceleration efficiency, The Astrophysical Journal 783~(2)
  (2014) 91.
\newblock \href {https://doi.org/10.1088/0004-637X/783/2/91}
  {\path{doi:10.1088/0004-637X/783/2/91}}.

\bibitem{Ha:2018}
J.-H. Ha, D.~Ryu, H.~Kang, A.~J.~v. Marle, Proton acceleration in weak
  quasi-parallel intracluster shocks: Injection and early acceleration, The
  Astrophysical Journal 864~(2) (2018) 105.
\newblock \href {https://doi.org/10.3847/1538-4357/aad634}
  {\path{doi:10.3847/1538-4357/aad634}}.

\bibitem{Ha2026}
J.-H. Ha, K.~Stasiewicz, Ballistic surfing acceleration as a coherent mechanism
  for electron acceleration in galaxy cluster shocks, MNRAS 550 (2026)
  stag1188.
\newblock \href {https://doi.org/10.1093/mnras/stag1188}
  {\path{doi:10.1093/mnras/stag1188}}.

\bibitem{Hoeft2007}
M.~Hoeft, M.~Br{\"u}ggen, Radio signature of cosmological structure formation
  shocks, MNRAS 375~(1) (2007) 77--91.
\newblock \href {https://doi.org/10.1111/j.1365-2966.2006.11111.x}
  {\path{doi:10.1111/j.1365-2966.2006.11111.x}}.

\bibitem{Siemieniec-Ozieblo2011}
G.~Siemieniec-Ozieblo, B.~Pasternak, Nonthermal emission from the radio relic
  of the galaxy cluster a2256, Astronomische Nachrichten 332~(6) (2011)
  649--654.
\newblock \href {https://doi.org/10.1002/asna.201011552}
  {\path{doi:10.1002/asna.201011552}}.

\bibitem{Pinzke2013}
A.~{Pinzke}, S.~P. {Oh}, C.~{Pfrommer}, {Giant radio relics in galaxy clusters:
  reacceleration of fossil relativistic electrons?}, MNRAS 435~(2) (2013)
  1061--1082.
\newblock \href {https://doi.org/10.1093/mnras/stt1308}
  {\path{doi:10.1093/mnras/stt1308}}.

\bibitem{Stroe2016}
A.~Stroe, T.~Shimwell, C.~Rumsey, R.~van Weeren, M.~Kierdorf, J.~Donnert, T.~W.
  Jones, H.~J.~A. Röttgering, M.~Hoeft, C.~Rodríguez-Gonzálvez, J.~J.
  Harwood, R.~D.~E. Saunders, The widest frequency radio relic spectra:
  observations from 150 mhz to 30 ghz, MNRAS 455~(3) (2016) 2402--2416.
\newblock \href {https://doi.org/10.1093/mnras/stv2472}
  {\path{doi:10.1093/mnras/stv2472}}.

\bibitem{Fermi:1949}
E.~Fermi, On the origin of the cosmic radiation, Phys. Rev. 75 (1949)
  1169--1174.
\newblock \href {https://doi.org/10.1103/PhysRev.75.1169}
  {\path{doi:10.1103/PhysRev.75.1169}}.

\end{thebibliography}

\end{document}